\documentclass[twocolumn,aps,prl,groupedaddress,showpacs,showkeys]{revtex4}
\usepackage{graphicx}
\usepackage{amsmath}
\usepackage{bm}

\begin{document}
\title{Spike Onset Dynamics and Response Speed in Neuronal Populations}
\author{Wei Wei and Fred Wolf}
\affiliation{MPI for Dynamics and Self-Organization, Faculty of physics, Georg-August University, \\ and Bernstein center for computational neuroscience, 
G\"{o}ttingen, Germany}

\begin{abstract}
Recent studies of cortical neurons driven by fluctuating currents
revealed cutoff frequencies for action potential encoding of
several hundred \textrm{Hz}.
Theoretical studies of biophysical neuron models have predicted a much
lower cutoff frequency of the order of average firing rate or the
inverse membrane time constant.
The biophysical origin of the observed high cutoff frequencies is
thus not well understood. Here we introduce a neuron model with
dynamical action potential generation, in which the linear response
can be analytically calculated for uncorrelated synaptic noise. We
find that the cutoff frequencies increase to very large values when
the time scale of action potential initiation becomes short.
\end{abstract}



\maketitle

\pagenumbering{arabic} In the cerebral cortex of the brain
information is encoded in the action potential (AP) firing rates of a
large ensemble of nerve cells. Recent experiments have observed
a surprisingly high cutoff frequency for the action potential encoding of
cortical neurons driven by fluctuating input currents
\cite{kodgen,boucsein,higgs,tatjana}. In a seminal paper K\"{o}ndgen
\textit{et al.} showed that the transmission function of layer 5
pyramidal neurons for a noisy sinusoidal signal does not decay until
about 200 Hz \cite{kodgen}. Later experiments confirmed such high
cutoff frequencies for signals coded by both the mean current and
noise strength \cite{boucsein} and in other types of cortical
neurons \cite{higgs,tatjana}. For an early observation of fast
response see \cite{silberberg}. Previous theoretical studies of
biophysical neuron models, however, predicted cutoff frequencies of
the order of the average firing rate or the inverse membrane time
constant (below 20 Hz), much lower than the experimentally observed values
\cite{fourcaud,fourcaud2005,naundorf2005}. Thus, the origin of the
high cutoff frequencies found in cortical neurons is currently not
well understood. Numerical investigation of neuron models with
dynamical AP generation, like the exponential integrate-and-fire
(EIF) model or the generalized theta neurons, suggested that details
of AP generation can influence the dynamical response of neuronal
populations
\cite{fourcaud,fourcaud2005,naundorf2005,richardson2007}. What is
missing, however, is a transparent understanding of how and when the
population cutoff frequency can dissociate from the basic single
neuron timescale set by the mean firing rate and the time constant
of membrane potential relaxation.

In this work we present an analytically solvable model which
explicitly describes the dynamical AP initiation process. A neuron
initiates an AP if the membrane potential passes an unstable fixed
point, the voltage threshold. In the leaky integrate-and-fire (LIF)
model, for which the linear response is known analytically, the
unstable fixed point coincides with the absorbing boundary and a
spike is triggered immediately when the membrane potential reaches
this threshold \cite{brunel,lindner}. As a consequence, boundary
induced artifacts dominate the response for high signal frequencies
in the LIF model \cite{fourcaud,fourcaud2005,naundorf2005}. One
important advantage of our new model is that such boundary induced
artifacts can be separated out mathematically, isolating the
physically meaningful part of the response function. We first
present the linear response for both encoding paradigms with white
noise. We find that for a wide range of parameter settings the
cutoff frequency is directly proportional to the AP onset rapidness
for a noise coded signal. It therefore dissociates from the membrane
time constant and can become arbitrarily large. For the mean current coded
signal, however, the cutoff frequency is confined by the membrane
time constant in the white noise case. We show by numerical
simulation that this confinement can be
broken when a finite correlation time in the synaptic noise is taken
into account and high cutoff frequencies can be obtained for a
large AP onset rapidness. Interestingly, experiments showed that the
AP onset rapidness of cortical neurons is very large both \textit{in
vitro} and \textit{in vivo} \cite{naundorf2006,badel}, which may
thus explain the occurrence of high cutoff frequencies. Our results
provide a relationship between the spike onset dynamics and the
population cutoff frequency that can be directly tested in
physiological experiments.

\begin{figure}[tbp]
  \centering
  \includegraphics[width=0.45\textwidth]{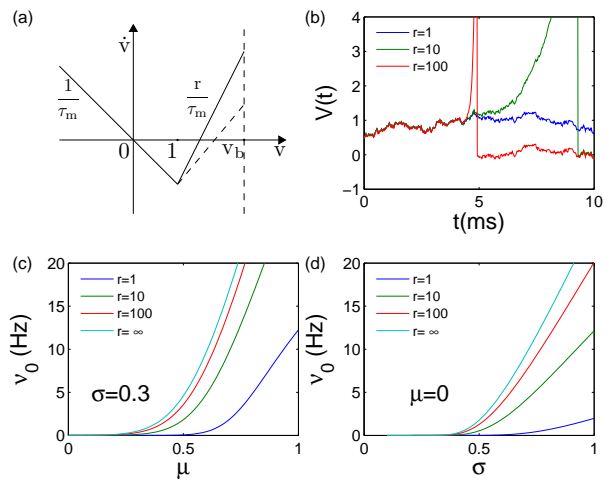}
  \caption{\label{fig1}(color online) (a) Illustration of the model. (b) $V(t)$
  trajectories for identical noise and three different values of r. (c) and (d) show
  the dependence of stationary firing rate on mean input current and noise strength
  in the noise driven regime. ($\tau_m=10\;\textrm{ms}$, $v_r=0$, $v_{b}=10$, $\tau_r=0$)}
\end{figure}

The simplest voltage dynamics that exhibits both a stable fixed
point (the resting potential) and an unstable fixed point (the
voltage threshold) has a piecewise linear membrane current, composed
of a leak current for low potential and a linear spike generating
current for high potential [Fig.~\ref{fig1}(a)]. The model is defined
by the following Langevin equation
\begin{equation}\label{lag1}
\tau_m\dot{v}=-v+\Theta(v-v_0)\; (r+1)(v-v_0)+\mu+\sigma \eta(t)\;,
\end{equation}
where $v$ is the membrane potential relative to the resting
potential, $\tau_m$ is the membrane time constant, $\Theta(v)$ is
the Heaviside step function, $r$ is the AP onset rapidness which
sets an effective time constant $\tau_m/r$ for the AP initiation
process. The larger is r, the faster is the spike onset
[Fig.~\ref{fig1}(b)]. In biophysical models, the onset rapidness will be set largely by 
intrinsic properties of the voltage dependent sodium channels, e.g., the
gating charge and the slope of the activation curve \cite{naundorf2006}.
$\mu$ is the mean input current and $\sigma$ is the amplitude of
synaptic noise. $\eta(t)$ is a Gaussian white noise satisfying
$\langle\eta(t)\rangle=0$ and
$\langle\eta(t)\eta(t^{'})\rangle=\tau_m\delta(t-t^{'})$.  The
crossing point $v_0$ of the two pieces sets the rheobase current,
which we use as the unit of voltage, $v_0=1$. The threshold
potential is $v_t=(1+1/r)v_0$.  When the membrane potential reaches
$v_b$, the truncation point of the AP upstroke, it is reset to a
voltage $v_r$ and stays there for an absolute refractory period
$\tau_r$. For convenience we take $\tau_m$ as the unit of time in
analytical calculation.

The Fokker-Plank equation corresponding to Eq. (\ref{lag1}) has the
following form
\begin{eqnarray}\label{fpe1}
&&\partial_t{P_1}+\partial_v\bigg(-v+\mu-\frac{1}{2}\sigma^2\partial_v\bigg)P_1= 0\;,\nonumber\\
&&\partial_t{P_2}+\partial_v\bigg(r(v-v_t)+\mu-\frac{1}{2}
\sigma^2\partial_v\bigg)P_2=0,\;
\end{eqnarray}
where $P_1(v,t)$ and $P_2(v,t)$ are the probability densities of
membrane potential $v$ for $-\infty< v\le 1$ and $1<v\le v_b$
respectively. The stationary firing rate can be found easily when
the boundary conditions are specified. From the reset assumption, we
impose an absorbing boundary at $v_b$ and, as a result the
probability density is zero there, $P_2(v_b,t)=0$. The firing rate
is given by the probability current at $v_b$, $\nu(t)=-\frac{1}{2}
\sigma^2\partial_vP_2(v_b,t)$. At the reset voltage $v_r$, the
probability density is continuous, while its first derivative has a
discontinuity from the reset condition: $\partial_v
P_1(v_r^+,t)-\partial_v P_1(v_r^-,t)=\partial_v P_2(v_b,t-\tau_r)$.
In addition, the density and its first derivative should be
continuous at $v=1$. When $\mu$ and $\sigma$ are constants, the
system is homogeneous and the stationary solution of Eq.
(\ref{fpe1}), denoted as $P_{01}(v)$ and $P_{02}(v)$ respectively,
can be found. The stationary firing rate $\nu_0$ is then obtained
from the normalization condition of the density \cite{sup}.

The stationary firing rate $\nu_0$ of the model Eq.
(\ref{lag1})reduces to that for the LIF model for $r\to \infty$.
Figure \ref{fig1}(c) and D show the dependence of $\nu_0$ on the mean
input $\mu$ and the amplitude of noise $\sigma$, respectively. The
firing rate $\nu_0$ increases monotonically with r, $\mu$, and
$\sigma$ and is relatively insensitive to $r$ when $r>10$. For the
dynamical response, however, the $r$ dependence is much more
pronounced. The instantaneous firing rate of an ensemble of model
neurons responds much faster for larger $r$ to a step change in the
noise level \citep{sup}.

\textit{Linear response.---} When the input current to a neuron is
weakly modulated, linear response theory can be applied to study the
dynamical response properties of an ensemble of neurons. To this
end, we consider a sinusoidal signal $\varepsilon \cos(\omega t)$,
where $\varepsilon$ is small. When the signal is encoded in the mean
current, $\mu\to \mu+\varepsilon \cos(\omega t)$ or in the noise
amplitude, $\sigma\to\sigma+\varepsilon\textrm{cos}(\omega t)$, the
instantaneous firing rate can be written as
$\nu(t)=\nu_0+\varepsilon|\nu_{1c}(\omega)|\textrm{cos}(\omega
t-\phi_c(\omega))$ or
$\nu(t)=\nu_0+\varepsilon\sigma|\nu_{1n}(\omega)|\textrm{cos}(\omega
t-\phi_n(\omega))$. Here $\nu_{1}(\omega)$ are the complex response
functions. The absolute value $|\nu_{1}(\omega)|$ are the
transmission functions and the phase angles
$\phi_1(\omega)=\arg(\nu_{1}(\omega))$ give the phase lags, which
completely characterize the linear response. Note that we refer to
both signal channels when the subscripts $c$ and $n$ are omitted.

It is known that the absorbing boundary condition at $v_b$ can
induce severe artifacts in the dynamical response. The potential
$v_b$ marks a "point of no return," which is not present in a
biophysical dynamical model of AP initiation. As a consequence, the
transmission function for a noise coded signal in the LIF model, for
instance, does not decay at high signal frequencies
\cite{silberberg,lindner}. Ideally one would thus wish to separate
the response function into a physiologically meaningful part
$\nu_1^{phy}(\omega)$ and a part containing all artifacts such that
$\nu_{1}^{phy}(\omega)=\nu_{1}(\omega)-\nu_1^{abs}(\omega)$.
$\nu_1^{phy}(\omega)$ must have the following properties: i)
$\nu_1^{phy}(\omega)$ approaches the static susceptibility when
$\omega\to 0$, specifically, $\nu_{1c}^{phy}(\omega)\to
\frac{\partial \nu_0}{\partial \mu}$ and $\nu_{1n}^{phy}(\omega)\to
\frac{1}{\sigma}\frac{\partial \nu_0}{\partial \sigma}$; ii)
$\nu_1^{phy}(\omega)\to 0$ when $\omega\to \infty$; iii) no
essential dependence on the truncation point $v_b$. The artifactual
part from the absorbing boundary should have the following
properties: i) negligible contribution for signal frequency in the
physiologically relevant range $f\le 1\textrm{kHz}$, e.g.
$|\nu_1^{abs}(\omega)|\ll |\nu_1^{phy}(\omega)|$, where
$f=\omega/2\pi$; ii) strong dependence on the truncation point
$v_b$. As we will show next such an isolation of the physiologically
meaningful response is possible in the model Eq. (\ref{lag1}).

\begin{figure}[tbp]
  \centering
  \includegraphics[width=0.42\textwidth]{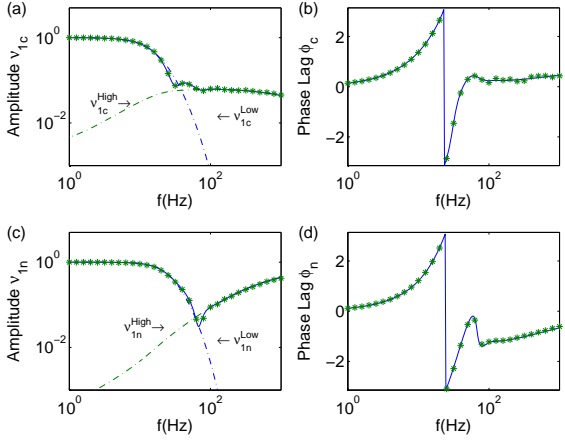}
  \caption{\label{fig2}(color online)
  The normalized function $\nu_{1}(\omega)/\nu_{1}(0.1)$
  and phase lag for a mean coded signal and noise coded signal with  $r=1$, $\mu=0$, and $\nu_0=5\;\textrm{Hz}$.
  Other parameters are the same as in Fig.~\ref{fig1}. Lines, theory; Symbols,
  simulation.}
\end{figure}

In the model Eq. (\ref{lag1}), the linear response can be obtained
analytically by expanding the probability density in Eq.
(\ref{fpe1}) to first order in $\varepsilon$ and using the Green's
function method. We find that $\nu_{1n}(\omega)$ decomposes
naturally into two parts,
$\nu_{1}(\omega)=\nu_{1}^{Low}(\omega)+\nu_{1}^{High}(\omega)$, with
\begin{widetext}
\begin{eqnarray}\label{nu1low}
\nu_{1c}^{Low}(\omega)&=&\frac{i\omega}{(1-i\omega)(1+i\omega/r)}
\frac{(1+1/r)(\psi_1P_{01}-\sqrt{D}\Phi_1P_{01}^{'})-
(1+i\omega/r)\Phi_1(v_r)e^{\Delta_0+i\omega\tau_r}}{
\psi_1(v_r)e^{\Delta_0+i\omega\tau_r}+(Y_1\psi_1^{'}-Y_1^{'}\psi_1)e^{\Delta_1}}\;,\nonumber\\
\nu_{1n}^{Low}(\omega)&=&\frac{i\omega(i\omega-1)}{(2-i\omega)(2+i\omega/r)}\;\frac{(1+1/r)
\big(\frac{i\omega}{(1-i\omega)\sqrt{D}} \Phi_1 P_{01} +2\Upsilon_1
P_{01}^{'}\big)
+\frac{\nu_0}{D}(2+i\omega/r)\Upsilon(v_r)e^{\Delta_0+i\omega\tau_r}}{
\psi_1(v_r)e^{\Delta_0+i\omega\tau_r}+(Y_1\psi_1^{'}-Y_1^{'}\psi_1)e^{\Delta_1}}\;,
\end{eqnarray}
\end{widetext}
where $D=\frac{1}{2}\sigma^2$, $\Delta_0=(1-v_r)(2\mu-1-v_r)/4D$ and
$\Delta_1=(1-v_b)(2\mu-1+r(v_b-v_t))/4D$. $\psi_1$, $\Phi_1$, and
$\Upsilon_1$ are parabolic cylinder functions, and $Y_1$, $Y_{2}$
are a combination of parabolic cylinder functions, whose definition
together with the $\nu_{1}^{High}(\omega)$ parts are given in the
supplement \citep{sup}. Any prime represents the derivative with
respect to $v$. Here and in the following, the functions adopt their
values at $v=1$ if not denoted explicitly. Figure ~\ref{fig2}
illustrates the linear response with $r=1$ as an example.

\textit{Removing boundary induced artifacts.---} The decomposition
of $\nu_{1}(\omega)$ into two additive components has exactly the
features required for the separation of artifacts. Using asymptotic
expansion of the parabolic cylinder functions we find that for a
finite signal frequency, $\nu_{1}^{High}$ can be approximated by
\begin{eqnarray}
\nu_{1c}^{High}(\omega)&\simeq&\frac{\nu_0}{r(v_b-v_t)+\mu}\frac{i\omega}{r+i\omega}\;,\nonumber\\
\nu_{1n}^{High}(\omega)&\simeq&-\frac{\nu_0}{(r(v_b-v_t)+\mu)^2}\;
\frac{i\omega(1+i\omega/r)}{2+i\omega/r}\;.\label{nu1nhigh}
\end{eqnarray}
when $v_b\gg v_{t}$. So $\nu_{1}^{High}(\omega)$ are strongly
dependent on $v_b$ and approach zero when $v_b\to \infty$ for finite
signal frequency. When $\omega\to 0$, $\nu_{1}^{High}(\omega)$ are
negligible compared with $\nu_{1}^{Low}(\omega)$ and are strongly
suppressed when $v_b$ is large. That $\nu_{1n}^{High}(\omega)$
captures all artificial contributions imposed by the absorbing
boundary condition is finally confirmed from the high frequency
behavior, $\nu_{1c}^{High}(\omega)\to
\frac{\nu_0}{\sqrt{D}}\frac{1}{\sqrt{\omega}}e^{i\pi/4}=\lim_{\omega\to
\infty} \nu_{1c}^{LIF}(\omega)$ and $\nu_{1n}^{High}(\omega)\to
\frac{\nu_0}{D}=\lim_{\omega\to \infty} \nu_{1n}^{LIF}(\omega)$,
since the high frequency behavior in the LIF model is determined
solely by the absorbing boundary.
As a consequence, neglecting $\nu_{1n}^{High}(\omega)$ in the
response function eliminates any boundary induced instantaneous
response components. These results establish that
$\nu_{1}^{Low}(\omega)$ capture the behavior of $\nu_{1}(\omega)$
for low and intermediate frequencies and decay to zero in the large
frequency limit. When $\omega\to 0$,
$\nu_{1n}^{Low}(\omega)\to\nu_{1n}(0)=\frac{1}{\sigma}\frac{\partial
\nu_0}{\partial \sigma}$, since $\nu_{1n}^{High}(\omega)$ is
negligible there. $\nu_{1}^{Low}(\omega)$ exhibits only a weak
dependence on $v_b$ through a frequency dependence phase lag
$\phi_0=\frac{\omega}{r}\log \frac{\dot{v}_b+\mu}{\sqrt{rD}}$,
characterizing the time lag to the truncation point $v_b$ of the AP
upstroke. Therefore, we have
$\nu_{1}^{Low}(\omega)=\nu_{1}^{phy}(\omega)$. The physiologically
meaningful predictions of the model can thus be revealed by
examining $\nu_{1}^{Low}(\omega)$ in isolation.

\begin{figure}[tbp]
  \centering
  \includegraphics[width=0.45\textwidth]{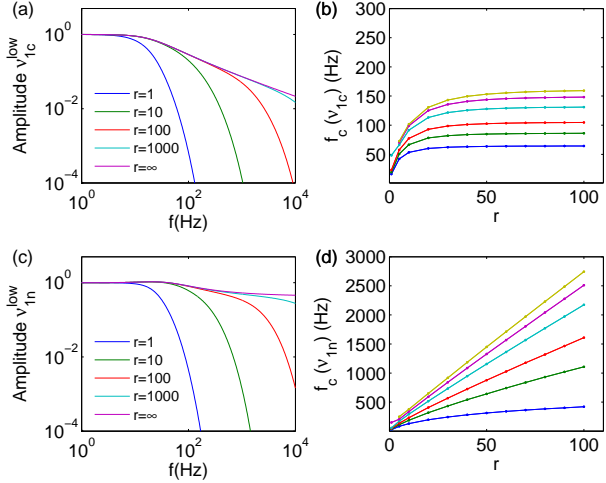}
  \caption{\label{fig4}(color online)
  (a) and (c) The normalized transmission function $\nu_{1}^{Low}(\omega)/\nu_{1}(0.1)$
  for different r with $\mu=0$, $\nu_0=5\;\textrm{Hz}$. 
  (b) and (d) The variation of cutoff frequency with $r$ for different
  firing rates: $\nu_0=1,5,10,20,30,40\;\textrm{Hz}$ from lower to upper curves. Other parameters are the same as in Fig. \ref{fig1}.}
\end{figure}

\textit{Cutoff frequency and AP onset rapidness.---}
Figure ~\ref{fig4} shows the behavior of $\nu_{1}^{Low}(\omega)$ with
increasing $r$ and how the cutoff frequency $f_c$ changes with r
for different $\nu_0$ \cite{expl}. We see that $f_c$ increases
linearly with the onset rapidness $r$ for noise coded signals when
the firing rate is not very low ($>1\;\textrm{Hz}$ here); while for
mean coded signals $f_c$ saturate for large $r$.
The increase of $f_c$ with firing rate $\nu_0$ results from the
stochastic double resonance phenomenon: the transmission function
will develop a peak for some optimal signal frequency before
decaying when $\nu_0$ is relatively large \cite{plesser}.

Figure ~\ref{fig4}D suggests that the cutoff frequency for a noise
coded signal follows $f_c\propto r$ and dissociates from $\tau_m$.
This is directly confirmed by the large frequency approximation of
$\nu_{1n}^{Low}(\omega)$, \small
\begin{eqnarray}\label{nu1lowapp}
\nu_{1n}^{Low}(\omega)&\propto&\frac{\exp(-\frac{\pi}{4}\omega/r)}
{2+i\omega/r}\bigg(-P_{01}^{'}+\frac{i\sqrt{i\omega/r}}{2\sqrt{D}}\tilde{P}_{01}\bigg)\;,
\end{eqnarray}
\normalsize where $\tilde{P}_{01}(v)\equiv\sqrt{r}P_{01}(v)$.
Because $\tilde{P}_{01}(v_0)\to
\sqrt{\frac{\pi}{2}}\frac{\nu_0}{\sqrt{D}}$ and $P^{'}_{01}(v_0)\to
-\frac{\nu_0}{D}$ for $r\gg 1$, the decay of
$\nu_{1n}^{Low}(\omega)$ depends essentially only on $\omega/r$.
This implies that the cutoff frequency for a noise coded signal
dissociates from $\tau_m$ and becomes proportional to the onset
rapidness $r$, $f_c=Ar$, where $A$ depends on $\nu_0$ through the
effect discussed above. This demonstrates that fast onset APs can
enhance the cutoff frequency and therefore the response speed
significantly. Note that a linear relationship $f_c\propto r$ was
previously conjectured by Naundorf \textit{et al.} based on
dimensional analysis \cite{naundorf2007}.

\begin{figure}[tbp]
  \centering
  \includegraphics[width=0.45\textwidth]{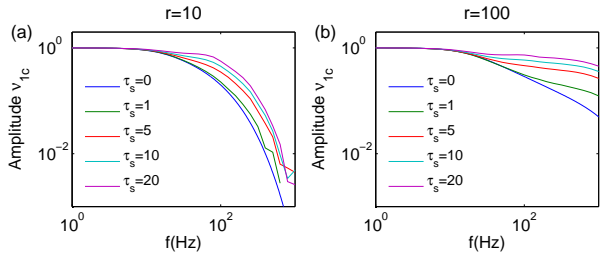}
  \caption{\label{nu1c-r-taus-fig}
  Variation of the normalized transmission function $|\nu_{1c}(\omega)/\nu_{1c}(0.1)|$ with
increasing correlation time $\tau_s$ (with unit $\textrm{ms}$) of the synaptic noise
for $r=10, 100$. Parameter used: $\mu=0$, $v_b=10$, $\nu_0=5
\textrm{Hz}$. }
\end{figure}

For a current coded signal, however, $\nu_{1c}^{Low}(\omega)\propto
\frac{1}{\sqrt{\omega}}\exp(-\frac{\pi}{4}\omega/r)$ for large $r$.
Therefore the linear response is confined by the membrane time
constant in the white noise case, as seen also from Fig.
\ref{fig4}A. Real synaptic inputs, however, have a finite
correlation time and can be approximated better as a colored noise.
As shown in Fig. \ref{nu1c-r-taus-fig}, the confinement by the
membrane time constant is broken under such conditions and cutoff
frequencies of several hundred Hz can be reached for large $r$
\cite{sup}.

Our results identify AP onset rapidness as a critical determinant of
population cutoff frequency and reveal how this cutoff frequency
can dissociate from the basic single neuron time constants set by
the mean firing rate and membrane time constant. The confinement of
the mean response for white noise constitutes an interesting
prediction of the model, which should be tested experimentally by
using a very short correlation time ($\leq 1 \;\textrm{ms}$) of the
synaptic noise. The origin of the large onset rapidness seen in
cortical neurons is a matter of ongoing debate \cite{mccormick,
naundorf2007}. Its value can be modified in real neurons by applying
drugs like TTX or knockout of sodium channel subtypes
\cite{naundorf2006,royeck}. Measurements of dynamical response for
cortical neurons under such manipulations are thus predicted to
provide important insight into the mechanism of fast population
coding.

We thank D. Battaglia and B. Lindner for suggestions and carefully
reading an early version of the manuscript. We are grateful to T. Geisel, M.
Gutnick, M. Huang, M. Monteforte, T. Moser and T. Tchumatchenko for
discussions. This work was supported by BMBF (01GQ07113),
GIF (906-17.1/2006), BCCN II (01GQ1005B), and DFG
(SFD899).


\begin{thebibliography}{99}
\bibitem{expl}For LIF model, $|\nu_{1n}(\omega)|$ starts at zero signal
frequency with a value $\frac{\partial \nu_0}{\partial D}$ and
saturates to a constant $\frac{\nu_0}{D}$ at large frequency. So in
our model there will be a transition between these two plateaus
before the decaying of transmission function, since in large $r$
limit it reduces to the LIF model. To characterize the decaying
property and avoid the transition region, we extract the cut-off
frequencies at which the normalized transmission function decays to
$1/\sqrt{10}$, which is smaller than the value usually adopted.
\end{thebibliography}
\end{document}